\documentclass[journal]{vgtc} 
\PassOptionsToPackage{dvipsnames,svgnames,table}{xcolor}
\usepackage{algorithm}
\usepackage{array}
\usepackage[caption=false,font=normalsize,labelfont=sf,textfont=sf]{subfig}
\usepackage{textcomp}
\usepackage{stfloats}
\usepackage{url}
\usepackage{verbatim}
\usepackage{graphicx}
\usepackage{cite}
\usepackage{pifont}
\usepackage{amsmath,amsthm,amssymb,amsfonts}

\usepackage{algpseudocode}
\usepackage{appendix}

\graphicspath{{figures/}{pictures/}{images/}{./}} 

\usepackage{tabu}                      
\usepackage{booktabs}                  
\usepackage{lipsum}                    
\usepackage{mwe}                       
\usepackage{mathptmx}                  
\usepackage{tikz}
\DeclareRobustCommand*\circled[1]{\tikz[baseline=(char.base)]{
            \node[shape=circle,draw,inner sep=1.1pt] (char) {#1};}}

\onlineid{1678}

\vgtccategory{Research}

\vgtcpapertype{please specify}

\title{MetaRoundWorm: A Virtual Reality Escape Room Game for Learning the Lifecycle and Immune Response to Parasitic Infections}

\author{%
  \authororcid{Xuanru Cheng}{0009-0003-9470-5537},
  \authororcid{Xian Wang}{0000-0003-1023-636X},
  \authororcid{Chi-lok Tai}{} and 
  \authororcid{Lik-Hang Lee}{0000-0003-1361-1612}
}

\authorfooter{
  \item
  	Xuanru Cheng is with the Hong Kong Polytechnic University.
  	E-mail: jiabao.cheng@connect.polyu.hk.
  \item
  	Xian Wang is with the Hong Kong Polytechnic University.
  	E-mail: xiann.wang@connect.polyu.hk.

    \item 
        Chi-lok Tai is with the Hong Kong Polytechnic University.
  	E-mail: andy.tai@cpce-polyu.edu.hk.
    
    \item 
        Lik-Hang Lee is with the Hong Kong Polytechnic University.
  	E-mail: lik-hang.lee@polyu.edu.hk (corresponding author).
    
}

\abstract{
   Promoting public health is challenging owing to its abstract nature, and individuals may be apprehensive about confronting it. Recently, there has been an increasing interest in using the metaverse and gamification as novel educational techniques to improve learning experiences related to the immune system. Thus, we present \textit{MetaRoundWorm}, an immersive virtual reality (VR) escape room game designed to enhance the understanding of parasitic infections and host immune responses through interactive, gamified learning. The application simulates the lifecycle of Ascaris lumbricoides and corresponding immunological mechanisms across anatomically accurate environments within the human body. Integrating serious game mechanics with embodied learning principles, \textit{MetaRoundWorm} offers players a task-driven experience combining exploration, puzzle-solving, and immune system simulation. To evaluate the educational efficacy and user engagement, we conducted a controlled study comparing \textit{MetaRoundWorm} against a traditional approach, i.e., interactive slides. Results indicate that \textit{MetaRoundWorm} significantly improves immediate learning outcomes, cognitive engagement, and emotional experience, while maintaining knowledge retention over time. Our findings suggest that immersive VR gamification holds promise as an effective pedagogical tool for communicating complex biomedical concepts and advancing digital health education. 
} 

\keywords{Virtual Reality, Game-Based Learning, Parasite Lifecycle Education, Gamification, MetaRoundWorm}

\teaser{
  \centering
  \includegraphics[width=\linewidth]{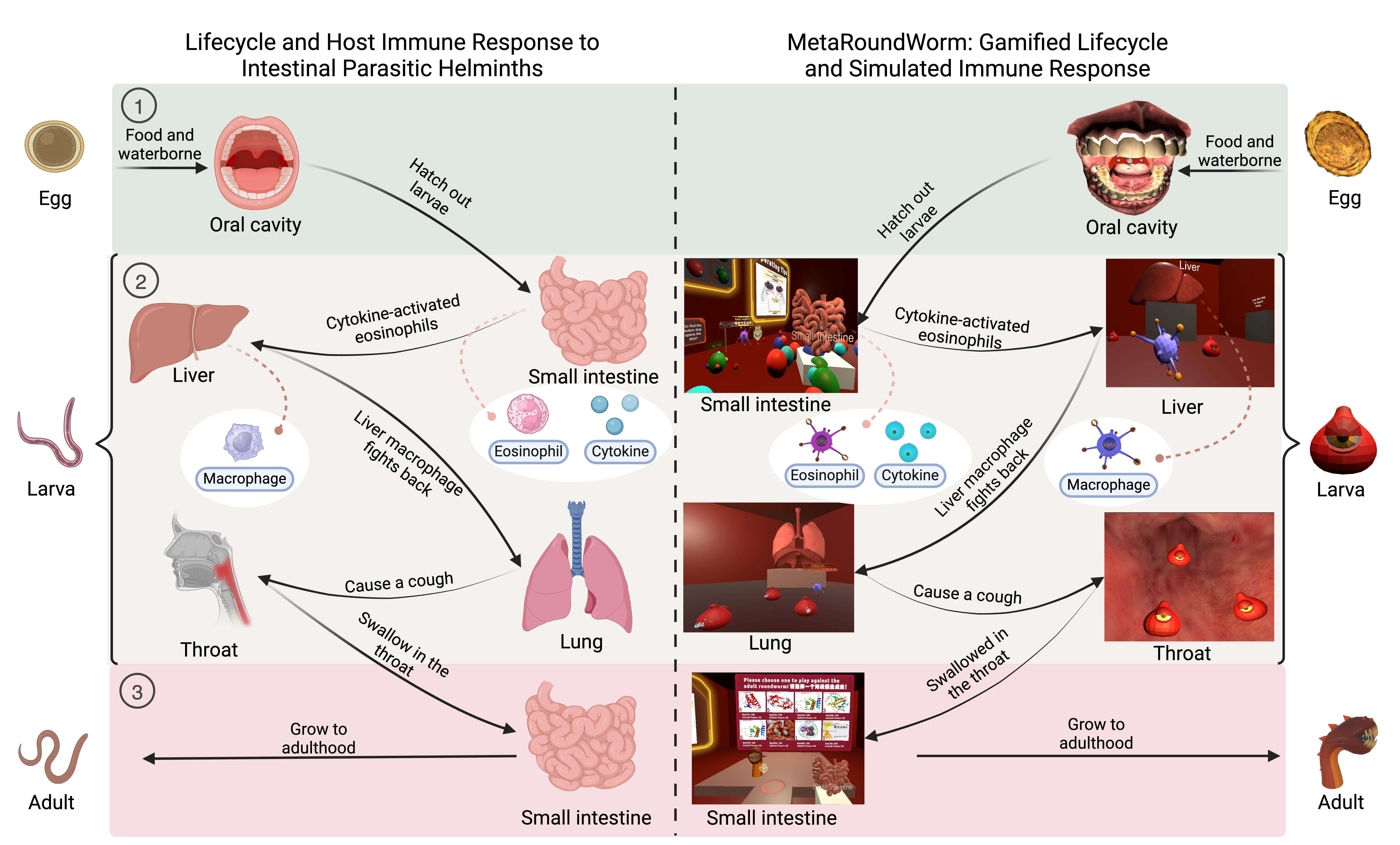}
  \caption{A comparative visualization of the biological and gamified lifecycle of intestinal parasitic helminths. (\textit{Left}) The natural lifecycle and host immune response to intestinal parasitic helminths, highlighting key anatomical and immunological interactions across three developmental stages: \circled{1} oral ingestion and intestinal hatching, \circled{2} larval migration and immune evasion, and \circled{3} maturation to adulthood in the small intestine. (\textit{Right}) Our VR escape room game, \textbf{MetaRoundWorm}, simulates this lifecycle by immersing players in a gamified representation of the same physiological and immunological processes. The game promotes learning through interactive challenges reflecting cytokine signaling, macrophage defense, and parasite-driven pathology.}
  \label{fig:teaser}
}

\begin{document}

\maketitle

\section{Introduction}



The immune system is an organization of cells and molecules with a special role in defense against infection~\cite{delves2000immune} and is the intrinsic essence of health~\cite{brodin2017human}. Parasites date back thousands of years, and parasitic infections have affected people's health for centuries~\cite{cox2002history}. Parasitic infections are widespread in the tropics and subtropics, and their immunopathology is of paramount importance~\cite{cohen1976immunology}. Parasitic infections are highly coexistent with malaria and HIV/AIDS in developing countries, often leading to co-infections, exacerbating anemia, increasing susceptibility to pathogen transmission and infection, and accelerating disease progression, yet unfortunately, this is an often-neglected tropical disease~\cite{hotez2008helminth}. Ascariasis lumbricoides is the most prevalent and largest of the human helminths; it is found throughout the world and is a cosmopolitan parasite~\cite{de2017ascariasis}. Children, especially those with weak awareness of public health, are often infected through contact with contaminated soil, fingers or toys, and the risk is higher in areas where human waste is used as fertilizer~\cite{DOLD2011632}. 

However, current educational methods may not effectively communicate the complexity of immune system disorders, resulting in people failing to realize the important role of the immune system in their health~\cite{10331286}. Currently, the primary means of raising awareness of diseases related to the immune system is through traditional textbooks and lectures, which are difficult to attract the interest of those being taught and do not allow students to learn deeply and retain their knowledge for an extended period of time~\cite{7014255}. These traditional educational methods lack interactivity, and although current traditional teaching has attempted to introduce interactive PowerPoints to add interest to the teaching~\cite{hastutik2022development}, they still fail to explain the dynamic nature of the immune system visually. Therefore, innovative educational methods are essential to improve learning in this field.

Game-based learning (or GBL) refers to environments that promote the acquisition of knowledge and skills through game content and activities, where problem solving and challenge mechanics provide a sense of achievement for the player/learner~\cite{QIAN201650}.GBL is an educational method that utilizes games to improve comprehension and memory, and it has been used very successfully in formal education, especially in military, medical, and sports training~\cite{pivec2003aspects}. Game-based learning helps to stimulate students' intrinsic motivation to learn and increase their attention span while providing them with a powerful sense of immersion~\cite{jaaska2022game}. Escape rooms—puzzle-based games where players must solve a series of interconnected challenges to "escape" from themed environments—have shown significant potential in education~\cite{makri2021digital}. Digital educational escape rooms (DEERs) have moved toward digitalization and demonstrated success in knowledge acquisition~\cite{mti6020008,makri2021digital}.

VR is a technology that has received much attention in recent years and is widely used in serious games, gamification, and game-based learning~\cite{ha2016vr}. Regarding user participation in the environment and narrative, it provides users with a positive and fun experience with significant potential for education~\cite{freina2015literature}. Immersive VR educational games integrate visual, auditory, plot, interactivity, and game mechanics to provide a tremendous new visual education way to explain and simulate abstract immunology knowledge, which can help students understand the process of parasites infecting the human immune system in a more in-depth way and break the limitations of traditional lecture teaching, which is boring and lacks practice opportunities~\cite{zhao2021comparison}. 

This paper introduces MetaRoundWorm, which provides an intuitive way of visualizing immunology's complex and abstract knowledge. The system serves to enhance learners' knowledge of the relationship between the immune system and parasitic infections. While existing VR biology educational games like InCell VR~\cite{nasharuddin2021incell} and The Body VR~\footnote{\url{https://store.steampowered.com/app/451980/The_Body_VR_Journey_Inside_a_Cell/}} excel at visualizing cellular and anatomical structures, they lack focus on dynamic pathogen-host interactions and immune responses. MetaRoundWorm addresses this gap by combining parasite lifecycle simulation with interactive immune system mechanics across multiple organ systems. Our system is characterized by VR and DEERs-based learning. Our design of study considers the traditional approach, using interactive slides and the MetaRoundWorm system. In our user study, we recruited 48 participants and evaluated the learning outcomes and learner engagement with the learning conditions as mentioned above. We measured key metrics, including knowledge acquisition and retention, user perceived workload, game experiences, and motion sickness. Our results highlight that MetaRoundWorm outperforms the interactive slides in both learning effectiveness and user perception.
The paper contributes to the application design and comprehensive evaluation in the domain of public health education with a specific focus on the immune system. Our findings provide design implications for future immersive VR gamification education, potentially applicable to other educational applications beyond public health. 

\section{Related Work}


\subsection{Virtual Reality-based Training \& Education}
Virtual reality is a technology that has attracted considerable interest in recent years due to its capacity to provide enjoyable and immersive experiences, functioning as an effective multimedia for game-based education~\cite{ha2016vr}.
VR-based game-based learning offers several advantages, including increased user engagement, as evidenced by the study through~\cite{gao2018learning}; and increased user knowledge retention, as the researchers found by comparing an immersive HMD-based aviation safety education game with traditional aviation knowledge card-based education~\cite{7014255}; In addition to teaching students, VR can also train teachers to improve their expertise~\cite{10316424}. VR-based learning has also been effective in sports training, with studies showing that VR has significantly improved the training of table tennis players~\cite{10765415}. And game-based VR sports training can address some of the challenges posed in public environments, improving overall health and social acceptance~\cite{10765408}. VR-based educational training is also being used in industry, where researchers are looking at possibly introducing biophilic elements into training scenarios to improve worker mental health and human potential~\cite{9995072}. In paediatric nursing, researchers have developed a VR-based immersive nursing education system to train nurse trainees in identifying and appropriately managing the emotional reactions of patients' parents~\cite{9995550}. For certain challenging domains (e.g., Explosive disposal~\cite{9995201}) and the domain studied in this article, VR training provides significantly superior results in assessment compared to real-world training. While these studies demonstrate VR's effectiveness across diverse educational domains, applying VR to teach complex immunological processes remains underexplored. Our work addresses this gap by leveraging immersive VR to visualize microscopic parasite-host interactions and immune responses.

\subsection{Immune System and Public Health}

The immune system protects the body from infections through a complex network of cells, organs, proteins, and tissues that work together to defend against pathogens~\cite{delves2000immune}. Immunity includes both innate and adaptive immunity, which interact to protect the organism. Innate immunity provides immediate, non-specific defense through physical barriers and immune cells (neutrophils, monocytes, macrophages, complement, cytokines, and acute phase proteins), whereas adaptive immunity, which is unique to higher animals, responds specifically to antigens through T-lymphocytes and B-lymphocytes~\cite{parkin2001overview}. It is increasingly recognized that parasitic infections, such as roundworms, have a wide-ranging impact on the host immune system and may affect concurrent infectious diseases such as malaria~\cite{holland2009predisposition, mwangi2006malaria}. Chronic ascariasis is known to cause developmental delays and cognitive impairment in children, while migration of roundworms in the liver and lungs can lead to pathologies, including Loeffler's syndrome. Acute ascariasis, on the other hand, is due to severe infections with high roundworm loads, which may lead to intestinal obstruction, serious complications and, in extreme cases, even death~\cite{holland2009predisposition}. 

Ascariasis, caused by infection with Ascaris lumbricoides, has seriously affected people's health for centuries~\cite{de2017ascariasis}. It is often found in developing countries, with major infections occurring in sub-Saharan Africa, the Americas, China and East Asia~\cite{DOLD2011632}. It is transmitted by ingestion of contaminated water, food or soil~\cite{hadush2016ascariasis}. Ascariasis has the potential to become massively endemic in these areas because people in these regions are often unaware of the roundworm infection's impact on the human immune system. The current educational approach to Ascaris lumbricoides primarily involves teachers' seminars, posters, comic books, puppet shows, drawing activities~\cite{al2014developing}, and cartoons~\cite{bieri2013development}. Instead, most approaches involve educating teachers about ascariasis, which is then taught to students by teachers~\cite{gyorkos2013impact}.  The difficulty of drawing attention to this easily overlooked disease with traditional teaching methods leads to more and more infected people~\cite{hobbs2019effects}. Therefore, there is an urgent need for an innovative educational approach to enhance understanding of parasitic infections associated with immune system diseases. Our article evaluates two interactive learning media, interactive slides and VR, which reinforce our understanding of designing educational immersive systems related to the immune system. 

\subsection{Game-Based Learning (GBL) Education}
Studies exploring the effectiveness of gamification commonly conflate it with game-based learning and serious games~\cite{caponetto2014gamification}. Gamification refers to the `use of game design elements in non-game environments'~\cite{deterding2011game} or ``the use of game-based mechanics, aesthetics, and game thinking to engage people, inspire action, facilitate learning and problem solving"~\cite{parapanos2021gamification}. Djaouti et al. define the serious game as an information-based application whose basic intention (not exclusive) is to combine teaching, learning, communication, and information sharing with the entertainment aspects of video games in a coherent way~\cite{djaouti2011origins}. Serious games focus on education (in all its forms) rather than entertainment~\cite{michel2013cognitive}. However, `Game-Based Learning' (GBL series) refers to the appropriate use of serious games in the learning process~\cite{roedavan2021serious}. 

Game-based learning has been widely used in the education industry as a new type of education~\cite{anastasiadis2018digital}. It is mainly used to teach abstract knowledge to students through games. Tang poetry is challenging for preschool children to understand. Researchers developed a digital game-based Tang poetry learning system, and the study found that the poetry learning game has a better immediate memorization effect than traditional teaching~\cite{7840236}. Game-based learning has also been used in first aid education for people with autism spectrum disorders (ASD), making educational activities more manageable and more motivating for them~\cite{6000343}. 
Escape room games have gained particular attention as an immersive and interactive form of game-based learning. Although MetaRoundWorm shares the key features of escape rooms, our work uniquely presents a multi-level immersive and interactive experience that enables learners to comprehend the roundworm lifecycle and its implications for human health. 

MetaRoundWorm advances the intersection of VR-based education, public health, and game-based learning by introducing the first immersive escape room game that visualizes microscopic parasite-host interactions and immune responses. Unlike existing educational approaches that rely on static materials or abstract representations, our work enables learners to experientially navigate through anatomically accurate human organs while actively participating in immunological processes, establishing a novel paradigm for teaching complex biomedical concepts through embodied VR interactions.
\section{MetaRoundWorm: VR Escape Room}

We designed and implemented an escape room serious game in VR, namely \textit{MetaRoundWorm}. To achieve the mutual context of entertainment, interactivity, and education, the game levels were designed with popular mini-games that can be combined with escape rooms, including shooting games, board games, puzzle games, and collecting games~\cite{charlotte2022most, qaffas2020operational}. The VR game allows users to experience and learn the process of roundworm infection in various human body organs. Through the immersive journey, the user can understand how the human immune system reacts to it. It is worth mentioning that the knowledge delivery in the game leverages scientific facts. As such, the immersive game journey presents various organs within a human body, such as in the game scenes, and highlights how the roundworm infection of the human body will pass through the organs, mainly the oral cavity, pharynx, lungs, small intestines, liver, etc. (See Figure~\ref{fig:teaser}).

\begin{figure*}[htb]
    \centering
    \includegraphics[width=\linewidth]{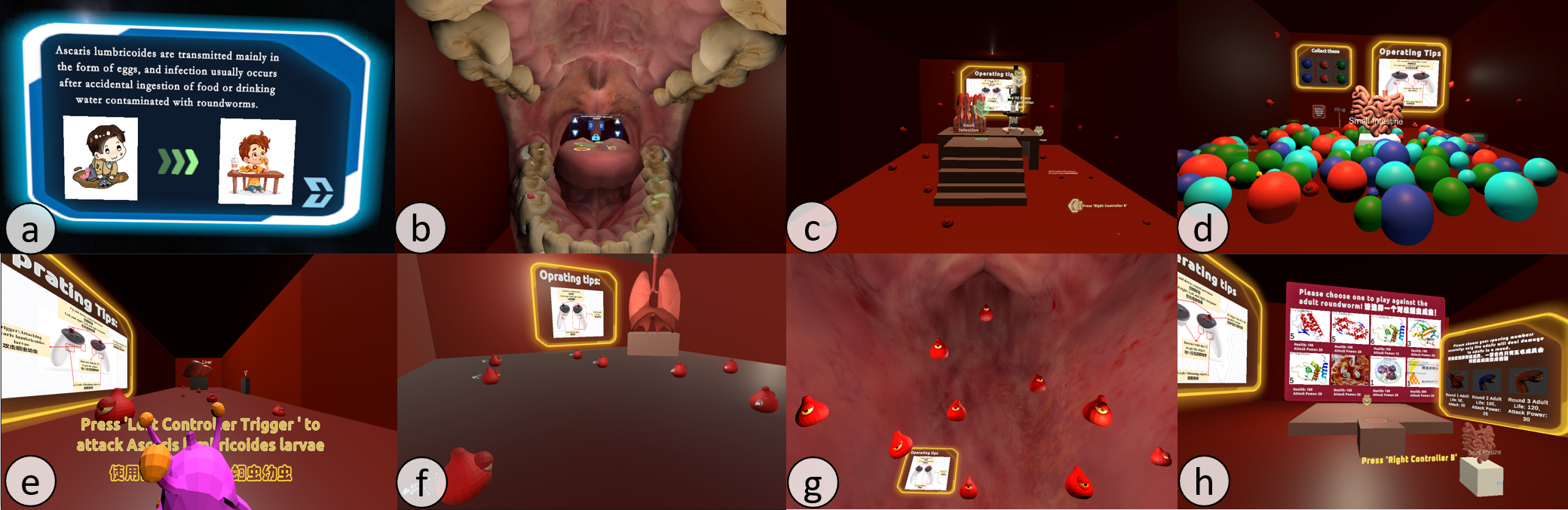}
    \caption{Gameplay Flow of VR Escape Room Game: a) [\textit{Prologue}] Introduction to the parasitic infection journey. b) [\textit{Oral Cavity}] Identify roundworm eggs hidden among food debris. c) [\textit{Small Intestine}] Trace the migration pathway of Ascaris lumbricoides. d) [\textit{Small Intestine}] Collect cytokines and activate immune cells. e) [\textit{Liver}] Deploy immune cells to eliminate migrating larvae. f) [\textit{Lung}] Simulate airflow to direct larvae movement during coughing. g) [\textit{Throat}] Expel roundworm larvae through a simulated cough reflex. h) [\textit{Small Intestine}] Eradicate adult roundworms using immune cells and antiparasitic drugs.}
    \label{fig:Game}
      \vspace{-5mm}
\end{figure*}


\subsection{Scientific Facts and Game Scenes}
Following the biology knowledge about roundworm infection in humans, Ascaris lumbricoides larvae first hatch in the small intestine and thereafter inhabit the liver and lungs throughout their larval stage~\cite{DOLD2011632}. First, the eggs are ingested and transported to the small intestine for hatching. The larvae subsequently migrate to the liver and enter the lungs within 6 to 8 days, where they penetrate the alveoli and ascend to the pharynx. Upon being swallowed, they return to the small intestine to mature into adult worms~\cite{DOLD2011632}. Accordingly, our VR gameplay consists of six sequential scenes that follow the life cycle of the roundworm, as follows: 
(1) oral cavity, including teeth, tongue and other elements; (2) small intestine; (3) liver; (4) lung; (5) throat; (6) small intestine; These scenes generally use red to present the internal structure of the body. Each level also presents the organ(s) at the respective body position to notify the game player about the game progress. At the same time, we put a model simulating Ascaris lumbricoides larvae to remind the player of the current Ascaris lumbricoides' lifecycle.

\subsection{User Interaction and Player View}

The game features a user-friendly navigation system with visual and audio components. Clear interface prompts help players by providing three types of assistance: instructions on how to use controls, explanations of educational content, and hints for solving puzzles. This helps players focus on learning about parasites without becoming frustrated by gameplay mechanics~\cite{charsky2010edutainment}. Additionally, carefully selected sound effects and background music create a more engaging experience, helping players feel immersed in the environment while also drawing attention to important elements. Research shows that combining visual information with appropriate audio enhances both player engagement and knowledge retention~\cite{moreno2007interactive}. 

Players with a first-person perspective (except the `Liver Room') serve as guardians of the body and actively support the role of immune system defenders by navigating the body's internal environment and completing the tasks (or puzzle). The first-person perspective enhances immersion~\cite{10.1145/2702123.2702256} while enabling players to comprehend the life cycle of roundworms across various organ systems. Player interactions include controls for navigation, object manipulation, and combat mechanics via the  controller's joysticks, buttons, and boards. These controls support core gameplay elements such as item collection, environmental puzzle-solving, and parasite elimination at different developmental stages.

\subsection{Game Levels and Education Content}
MetaRoundWorm is designed to achieve specific educational objectives aligned with Bloom's taxonomy~\cite{krathwohl2002revision}. Our learning framework covers the first four cognitive levels: (1) \textbf{\textit{Remember}}: players remember the stages of Ascaris lumbricoides lifecycle; (2) \textbf{\textit{Understand}}: players understand mechanisms of host immune responses including cytokine signaling and macrophage activation; (3) \textbf{\textit{Apply}}: players simulate the execution of immune cell defence behaviours; (4) \textbf{\textit{Analyze}}: players analyze the properties of different immune cells.
We describe the game levels, including the game genres, educational content, and goals, as follows. Figure~\ref{fig:Game} shows screenshots of each scene in the \textit{MetaRoundWorm} gameplay flow. 

\paragraph{\textit{Prologue}}

The game employs a standard navigational interface with a conventional start screen, allowing a player to reach the content through a selection button. To enhance educational engagement, we implemented a narrative prologue that utilizes immersive visual guides to demonstrate roundworm transmission vectors and human body infiltration mechanisms so that the player can \textbf{\textit{understand}} how Ascaris lumbricoides enter the human body. This instructional prologue serves dual pedagogical and experiential functions by contextualizing subsequent gameplay activities within established parasitological frameworksview~\cite{article}.

\paragraph{\textit{Level 1: The Oral Cavity Room}}

In the `Oral Cavity Room,' players search for tiny roundworm eggs hidden in food, which teaches them how people typically get infected with these parasites by accidentally eating contaminated food~\cite{bethony2006soil}. Players use their Meta Quest 3 controller to grab food items, which automatically brings up information screens showing what roundworm eggs look like and explaining essential details about them. When players release the grab button, the food item stays where they placed it, and the information screen disappears.
To help players better \textbf{\textit{remember}} what they have learnt, we created a counting challenge where players must find and count all the parasite eggs in the room. The total number becomes the password they need to advance to the next level. Prior research shows that actively engaging with information through tasks like observation and counting helps people remember better than just reading or listening ~\cite{chi2014icap}.

\paragraph{\textit{Level 2: The Small Intestine Room}}

The `Small Intestine Room' uses a three-phase learning approach combining observation with hands-on simulation. First, players can use a microscope station how roundworm larvae travel through different human organs. This optional feature mimics how real scientists work: observation sometimes is a choice, but missing information can affect people's understanding ~\cite{hmelo2007scaffolding}. When players engage with the microscope, they can see how the parasites typically move from the liver to the lungs~\cite{DOLD2011632}. Next, players collect important immune system proteins called cytokines using their controllers. They need to find six different cytokines hidden throughout the intestinal environment. This design helps players learn about immune components in their natural biological context, making abstract concepts more concrete~\cite{ squire2006content}. In the final stage, players these collected cytokines to activate special immune cells called eosinophils, which accurately show how the human body actually responds to roundworm infections.

\paragraph{\textit{Level 3: The Liver Room}}

In the `Liver Room,' players switch to a third-person view (seeing their character from outside) as they control an immune cell called a macrophage. This view helps players feel connected to their character while reducing motion sickness that can occur in VR~\cite{jerald2015vr}. The room shows how specialized liver immune cells interact with roundworm larvae as they pass through this organ~\cite{deslyper2019liver}. The gameplay is straightforward: the immune cell automatically locks onto nearby parasite targets, and players simply pull the controller's trigger to shoot attacks, representing how actual immune cells release chemicals to kill invaders.
This simplified version of complex immune processes makes the science easier to \textbf{\textit{understand}} while still teaching accurate concepts~\cite{kiili2005digital}. When players successfully destroy all the parasite larvae, they earn a key to progress to the next level, providing a sense of achievement while accurately following the parasite's journey through the body~\cite{jerald2015vr}. 

\paragraph{\textit{Level 4: The Lung Room}}

The `Lung Room' shows how the lungs expel roundworms through coughing, a natural defense mechanism when Ascaris larvae move through the respiratory system~\cite{sarinas1997ascariasis}. Players use an upward cranking motion with their controllers to simulate taking a deep breath before a cough. This physical movement helps players learn about body functions by related actions, as research on embodied learning games implies improved \textbf{\textit{understanding}}~\cite{10.1145/2851581.2892455}. The cranking motion facilitates players' comprehension of how lungs expel parasites towards the throat via a physical process. As such, players' actions move the parasites to specific areas in the lungs where they would naturally be expelled, creating a learning experience where physical movements have both educational value and advance the game's story~\cite{antle2013getting}.

\paragraph{\textit{Level 5: The Throat Room}}
The `Throat Room' presents how the human body defends itself when roundworm larvae reach the throat, specifically demonstrating the natural coughing reflex triggered by parasite presence~\cite{leles2012ascaris}. Players use their controllers to guide the larvae toward the upper throat area, while realistic coughing sounds play to help players \textbf{\textit{understand}} what is happening in a real infection. This combination of visual gameplay and sound helps players better \textbf{\textit{remember}} these concepts by engaging multiple senses~\cite{mayer2003nine}. When players successfully complete this level, they receive anti-parasite medication as a reward, teaching them about actual medical treatments while bringing the storyline of infection and treatment to a satisfying conclusion.


\paragraph{\textit{Level 6: Revisiting the Small Intestine Room}}
\label{sec: Level 6}
The culminating environment returns players to the small intestine, where Ascaris lumbricoides larvae mature into adults~\cite{DOLD2011632}. This chamber implements a turn-based combat simulation synthesizing previously encountered immunological concepts. Players interact with a hybrid interface combining 2D selection panels with 3D object manipulation~\cite{lindgren2016enhancing}, \textbf{\textit{analyzing}} and selecting appropriate immune effectors based on attack values, a mechanic that implements the principles of retrieval practice~\cite{karpicke2011retrieval}. Following selection, players grab and throw virtual immune components using controller gestures. The game presents three consecutive battles against adult worms with increasing health points, with each round limiting players to five effective immune tools. This design requires players to \textbf{\textit{apply}} the most appropriate immune responses, reinforcing their understanding of how specific immune mechanisms combat parasitic infections~\cite{maizels2016regulation}. Upon completion, players receive positive reinforcement visually through celebratory effects, implementing reward structures that enhance retention while providing narrative closure~\cite{HAMARI2016170}.

\subsection{Apparatus}

We developed the system using Unity (ver. 2022.3.17f1) on a laptop equipped with a 13th Gen Intel Core i9-13900HX processor, 32GB RAM, and an NVIDIA GeForce RTX 4060 GPU, running the operating system Windows 11. The participants used the Meta Quest 3 HMD, which supports a resolution of $2064 \times 2208$ pixels and a $120 Hz$ refresh rate. User interactions were implemented using the Meta Interaction SDK.

\section{Experimental Design}
To investigate the effectiveness of immersive VR-based learning compared to traditional digital instruction, we designed a between-subjects study with two study conditions. \textbf{(1) Interactive slides:} Participants in this condition engaged with a web-based interactive slideshow built using Pear Deck~\footnote{\url{https://www.peardeck.com/}}. The slides incorporated static visuals, annotated diagrams, multiple-choice quizzes, and short interactive elements that simulate classroom instruction (see examples in Figure~\ref{fig:slide}). This condition served as the baseline for a traditional and technologically advanced educational format commonly used in universities~\cite{haryani2021impact}. \textbf{(2) MetaRoundWorm:} Participants in this condition experienced our MetaRoundWorm VR escape room game using a VR headset. MetaRoundWorm guided learners through the lifecycle of Ascaris lumbricoides, simulating key immune responses via gamified tasks and embodied interactions within virtual environments. The condition emphasized immersive, narrative-driven, and spatial-contextualized learning.

We propose to evaluate the following hypotheses: \textbf{[H1]}: MetaRoundWorm participants will achieve significantly higher immediate learning outcomes than those using interactive slides~\cite{shi2022effect}; \textbf{[H2]}: MetaRoundWorm participants will demonstrate significantly better knowledge retention in follow-up assessments~\cite{7014255}; \textbf{[H3]}: MetaRoundWorm participants will report significantly more positive emotional states (POMS-SF)~\cite{pallavicini2020virtual}; \textbf{[H4]}: MetaRoundWorm participants will experience significantly better user experience (AttrakDiff, GEQ)~\cite{shelstad2017gaming}; \textbf{[H5]}: MetaRoundWorm will increase the subjective perceived workload of the participants (NASA-TLX)~\cite{tinco2024analysis}; \textbf{[H6]}: The participants' behaviors will be different under these two instructional interventions.



\subsection{Assessment Design}

To evaluate the learning effectiveness of each instructional condition, we administered three assessments: \textbf{(1) Pre-test}, \textbf{(2) Post-test}, and \textbf{(3) Follow-up test}. All tests were designed to measure knowledge acquisition, retention, and conceptual understanding of roundworm infections and associated immune responses. \textbf{Pre-test:} Administered immediately before the learning session, the pre-test served to assess the baseline knowledge of the participants about Ascaris lumbricoides infection, including its transmission pathways, life cycle, affected organs, immune mechanisms and available treatments. This ensured comparability between the two groups prior to exposure to their respective learning conditions. \textbf{Post-test:} Immediately following the learning session, participants completed the Post-test. This test measured immediate learning gains attributable to the instructional intervention. \textbf{Follow-up test:} One week after the intervention, participants completed a Follow-up test identical in content to the Post-test. This assessment was designed to evaluate long-term knowledge retention and determine the lasting impact of each teaching method.


All tests consisted of 10 multiple choice questions (see Table~\ref{Tab:Questions}) that were reviewed by an expert in the medical domain to ensure the validity of the content. Each assessment (pre-test, post-test, and follow-up test) evaluated the same conceptual knowledge but with varied question wording, order, and phrasing to minimize testing effects. The change in scores across the three test points enabled a comparative analysis of both immediate and sustained learning outcomes between interactive slides and MetaRoundWorm conditions.

\subsection{Procedure, Measurement, and Participants}

\begin{figure*}[htb]
    \centering
    \includegraphics[width=\linewidth]{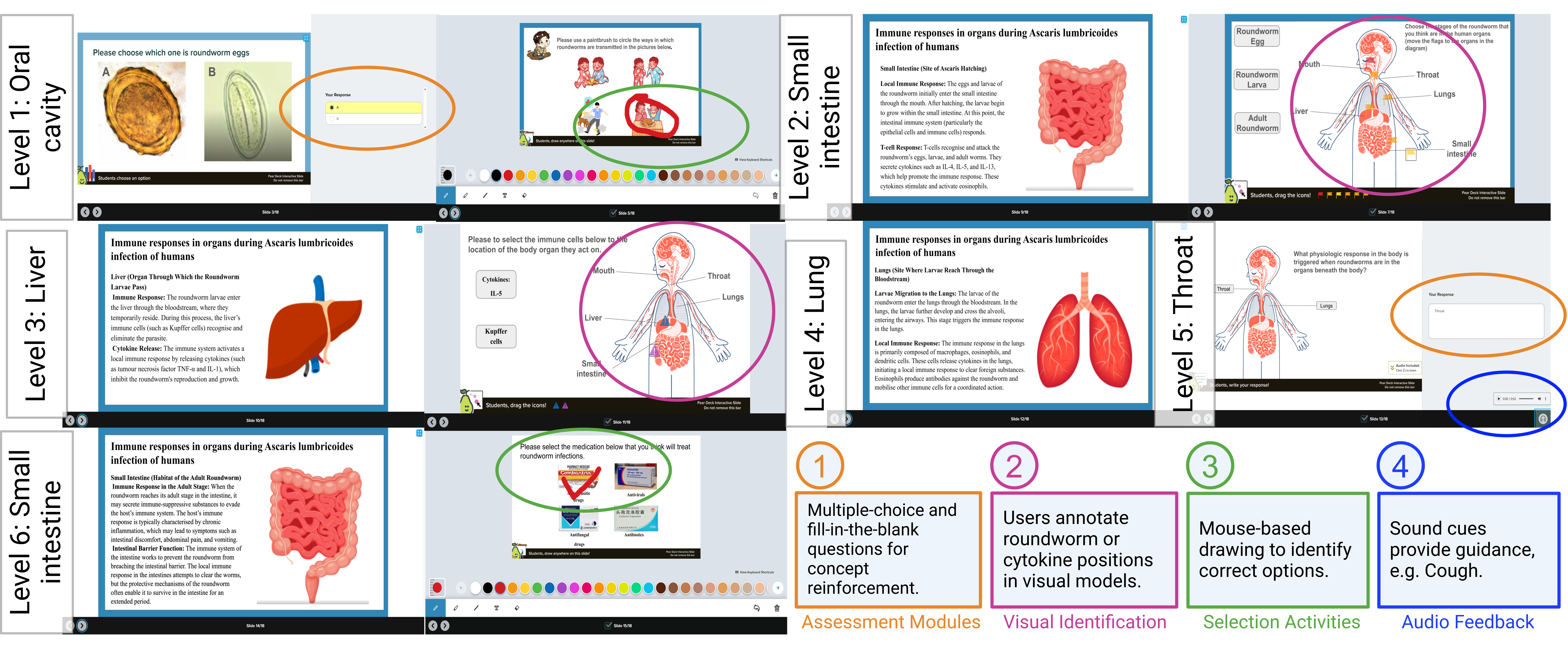}
    \caption{Interactive slides examples covering the same six levels as MetaRoundWorm. Each slide integrates four core interaction types: (1) assessment (multiple-choice and fill-in-the-blank questions); (2) visual identification (labeling roundworm or cytokine positions); (3) selection activities (mouse-based drawing tasks); and (4) audio feedback (contextual sound cues).}
      \vspace{-5mm}
    \label{fig:slide}
\end{figure*}

Before the experiment, the participants received a brief introduction to the experiment, provided informed consent, and completed a demographic questionnaire. A pre-test was administered to assess baseline knowledge and ensure comparability between groups. The participants were then randomly assigned to either the Interactive Slides group or the MetaRoundWorm group. Participants in the MetaRoundWorm group received a 10-minute orientation on using VR equipment, including navigation and object interaction. Once participants confirmed their familiarity with the controls of the VR system, the MetaRoundWorm educational experience began. To ensure comparable levels of engagement, both groups were given the same instructions: ‘Please complete the entire learning module at your own pace and make sure you understand each concept before moving on’. Both groups were given 60 minutes to work through the module, with the Interactive Slides group requiring interaction (answering questions, completing tasks) on each slide before moving on, ensuring full participation rather than simply skimming through the module, and the MetaRoundWorm group completing the tasks on each level before moving on to the next. Both educational methods used the same content reviewed by medical experts to ensure consistency. Each group consisted of 24 students and both groups were taught in similar controlled environments, differing only in the medium of instruction. After engaging with either the Interactive Slides or MetaRoundWorm, participants immediately completed the post-test, followed by four questionnaires to evaluate their subjective experiences. Finally, a brief interview was conducted to gather insights into participants' preferences, challenges encountered, and suggestions for improvements in both instructional methods. One week later, the participants returned to the experimental site to complete the follow-up test, evaluating long-term knowledge retention.

To evaluate user task performance and experience across the three conditions, we employed both quantitative and qualitative analyses. For objective measurements, we recorded participants' learning time and scores on three assessments, and during the experiment a researcher made informal observations of participants’ behaviour, including verbal communication, emotional expression and body movements. For subjective assessments, We administered the NASA-TLX Scale (NASA-TLX) to measure subjective perceived workload~\cite{hart1988development}, the AttrakDiff questionnaire to evaluate user experience with the interactive product~\cite{hassenzahl2003attrakdiff}, the Profile of Mood States - Short Form (POMS-SF) to assess short-term mood changes in participants~\cite{curran1995short}, and the Simulator Sickness Questionnaire (SSQ) to measure simulator and cybersickness~\cite{kennedy1993simulator}. We also used the Game Experience Questionnaire (GEQ) to evaluate the experiences and perceptions of the participants during gameplay~\cite{ijsselsteijn2013game}. Additionally, participants in the MetaRoundworm group completed the Simulator Sickness Questionnaire (SSQ) to assess any motion sickness or discomfort they might have experienced during the VR session~\cite{kennedy1993simulator}. To gain deeper insights into participants' learning experiences, we conducted semi-structured interviews immediately after each learning session. This qualitative approach complemented our quantitative findings by exploring participants' subjective experiences, learning strategies, and emotional responses.

\begin{table}[htb]
\caption{Assessment Questions List (Use Pre-test as an Example).}
\label{Tab:Questions}
\resizebox{1\linewidth}{!}{ 
\begin{tabular}{p{20pt}p{250pt}}
\toprule
Q1  & What is the main way roundworms infect the body?                                                          \\
Q2  & Which stage of the roundworm's life cycle in the human body passes through the lungs?                     \\
Q3  & In which part of the body do roundworms first hatch when they infect the body?                            \\
Q4  & Which medication is usually used in the treatment of roundworm infections?                                \\
Q5  & What are the usual characteristics of roundworm eggs?                                                     \\
Q6  & What physiological response in the body is triggered by roundworm larvae in the pharynx?                  \\
Q7  & Which substances activate macrophages during roundworm infection of the human immune system?              \\
Q8  & Which immune cells would be responsible for recognising and destroying larvae during roundworm infection? \\
Q9  & In which organ of the body do roundworms eventually mature?                                               \\
Q10 & Which organs of the body do roundworms pass through from egg -- larva -- adult?     \\ 
\bottomrule
\vspace{-5mm}
\end{tabular}}
\end{table}

We recruited 48 participants (25 female, 22 male and 1 prefer not to say) from local universities between the ages of 18 -- 40 years (M = 22.58, SD = 3.79). The familiarity of the participants with VR technology was measured on a 5-point scale, with 1 indicating no familiarity at all and 5 indicating a high level of familiarity, the mean VR familiarity score was 3.06 (SD = 9.92). All participants are graduate (41.67\%) and postgraduate (58.33\%) students. The research study was approved by the university's ethics review board.

\section{Results}
For all data analysis, we first tested normality with the Shapiro-Wilk test. In instances where the normality assumption was not met, we employed non-parametric alternatives, i.e., Mann-Whitney U tests, to appropriately analyze the data. For dimensions that meet parametric assumptions, independent sample t-tests were performed. 
\begin{figure}[htb]
  \centering
  \includegraphics[width=1\linewidth]{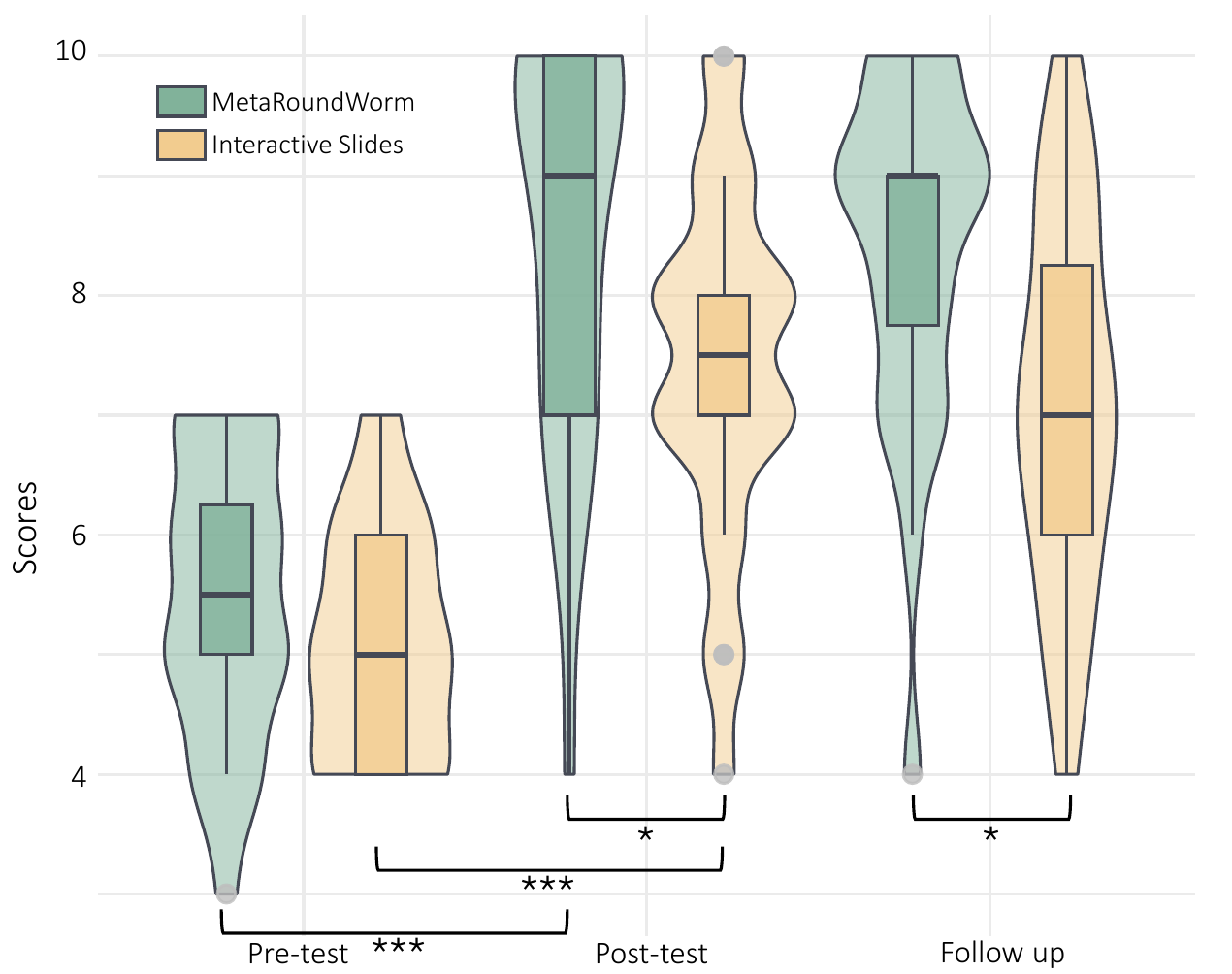}
  \caption{Violin and box plots of three assessments scores ($p < .05$(*), $p < .01$(**), $p < .001$(***)).}
  \vspace{-5mm}
  \label{fig:score}
\end{figure}

\subsection{Knowledge acquisition and retention}

Figure~\ref{fig:score} presents the distribution of the participant scores in the three assessments administered under two distinct instructional interventions. The data illustrate clear variations in performance, highlighting differential impacts of instructional methods on participants' outcomes. 
\paragraph{Immediate Learning Results}

We found that there was no significant difference between the pre-test scores of the MetaRoundWorm group and the Slide condition group ($U = 102$, $p = .175$, $r = .13$). This suggests that our two groups of participants had similar knowledge backgrounds and had similar perceptions of the topic of the roundworm prior to the instructional intervention. To evaluate immediate learning outcomes, post-test scores were compared between the two instructional conditions: MetaRoundWorm and Interactive Slides. Both groups exhibited significant gains from their respective pre-test scores (MetaRoundWorm: $t(15) = 7.24$, $p < .001$, $d = 1.81$; Interactive Slides: $t(14) = 4.72$, $p < .001$, $d = 1.22$), confirming that both interventions were effective in facilitating short-term learning. Notably, the MetaRoundWorm group achieved significantly higher post-test scores than the Interactive Slides group ($U = 138.5$, $p < .05$, $r = .35$), indicating a more robust immediate learning effect. Analysis of the violin plots further reveals a positively skewed score distribution for the MetaRoundWorm group at post-test, with a high concentration near the upper bound of the scale (scores of 9–10), suggesting a ceiling effect and strong mastery across participants. In contrast, the Interactive Slides group displayed a broader and more symmetric distribution centered around lower scores (approximately 7–8), indicating greater variability and lower overall performance.

 \paragraph{Knowledge retention effect}

Follow-up scores were used to assess knowledge retention over time. While both groups showed slight declines from their post-test levels, no significant differences were observed between post-test and follow-up test scores (MetaRoundWorm: $t(20) = -.96$, $p = .348$, $d = -.21$); Interactive Slides: $t(21) = -.82$, $p = .419$, $d = -.18$). However, the MetaRoundWorm group maintained significantly higher scores than the Interactive Slides group ($U = 133$, $p < .05$, $r = .37$). The MetaRoundWorm distribution remained tightly clustered around the upper score range, consistent with durable retention. In contrast, the Interactive Slides group exhibited a wider, flatter distribution, with scores dispersed across a broader range, suggesting less consistent retention. These distributional patterns underscore the greater efficacy of the MetaRoundWorm approach in promoting both immediate learning and long-term conceptual retention.

\subsection{Relevance analysis}
\begin{figure}[htb]
  \centering
  \includegraphics[width=1\linewidth]{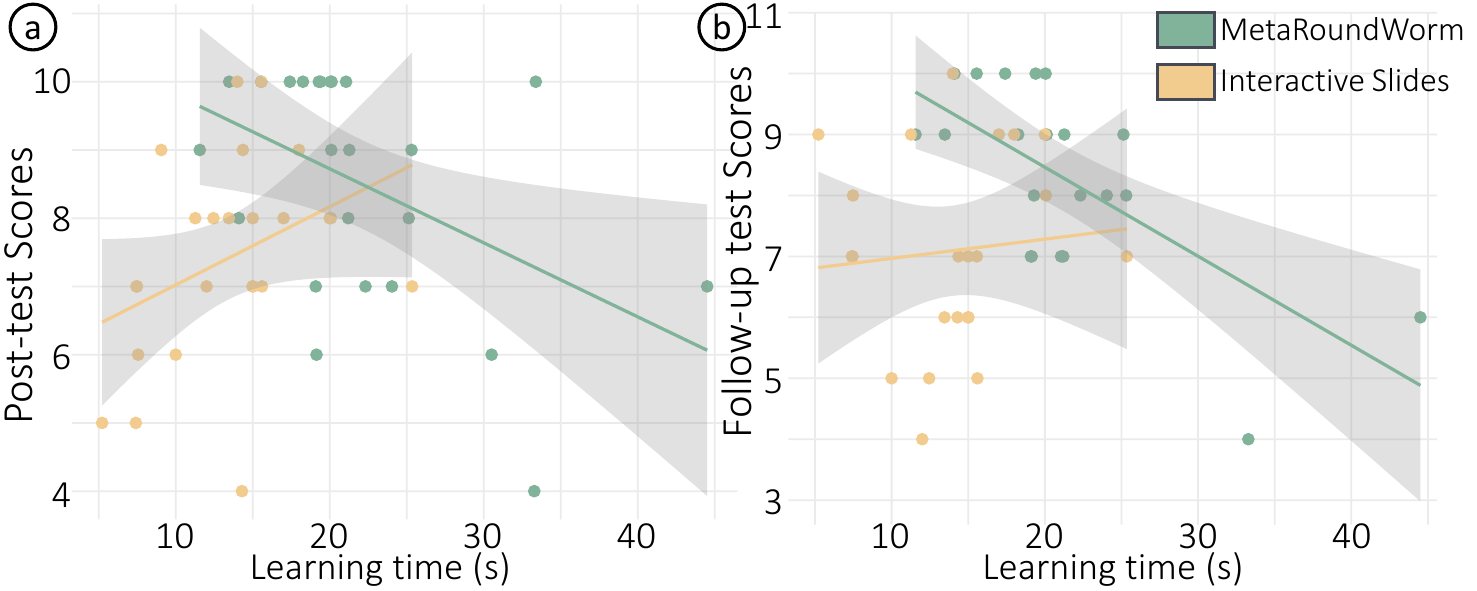}
  \caption{Relationship Between Learning Time and Assessment Outcomes with Linear
Regression. (a) Post-test scores vs. Learning time, (b)  Follow-up test scores vs. Learning time.} 
  \label{fig:linear}
    \vspace{-5mm}
\end{figure}

To assess the distribution of learning time and test scores, we conducted the Shapiro-Wilk tests. The results indicated significant deviations from normality for both score ($W = .91$, $p < .01$) and time ($W = .93$, $p < .01$), thus, we employed Spearman's rank-correlation analysis. At the general level, no significant correlation was observed between learning time and test scores among all participants ($\rho = .17$, $p = .237$, $R^2 = .03$, 95\% CI [-.12, .44]). However, subgroup analyzes revealed significant moderate correlations in opposite directions. Figure~\ref{fig:linear} displays the linear regression models fitted to the Post-test and Follow-up test scores over time. In the \textit{MetaRoundWorm}, learning time was negatively correlated with test performance ($\rho = -.41$, $p < .05$, $R^2 = .17$, 95\% CI [-.70, -.01]), suggesting that participants who spent less time in MetaRoundWorm tended to achieve higher scores. In contrast, a positive correlation was found in the \textit{Slide} condition ($\rho = .42$, $p = .041$, $R^2 = .18$, 95\% CI [.02, .70]), indicating that longer learning durations were associated with better performance. However, in follow-up test scores, the MetaRoundWorm group again exhibited a negative relationship between learning time and test performance ($\rho = -.55$, $p < .05$, $R^2 = .31$, 95\% CI [-.80, -.16]), while the Interactive Slides group no longer had significant positive associations ($\rho = .07$, $p = .65$, $R^2 = .02$, 95\% CI [-.30, -.53]). Notably, the dispersion of scores in the MetaRoundWorm condition was greater, particularly at higher learning times, as indicated by the wider confidence intervals. This may be related to the exploratory behavior of some participants (see Section~\ref{sec:Interactive}). These findings suggest differential effects of learning time on knowledge retention and immediate comprehension, depending on the instructional modality. While Interactive Slides may benefit from extended engagement, MetaRoundWorm performance appears to peak at lower durations of learning time.

\subsection{Subjective experience evaluation}

\begin{figure*}[htb]
  \centering
  \includegraphics[width=1\linewidth]{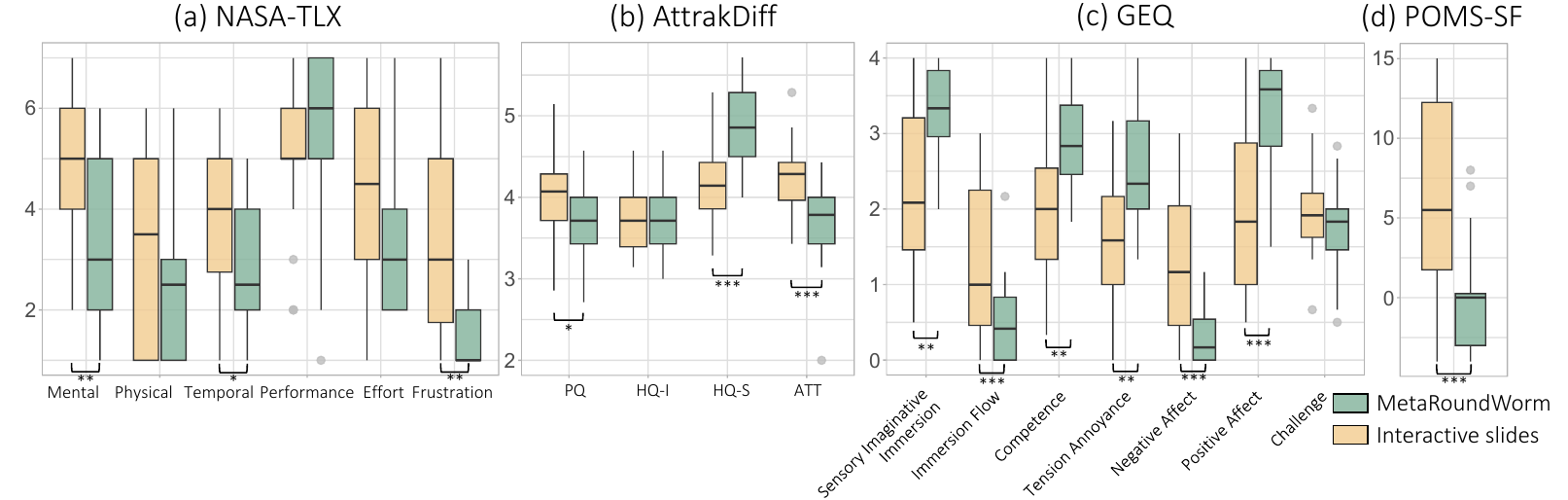}
  \caption{Box plots of subjective questionnaires for user study ($p < .05$(*), $p < .01$(**), $p < .001$(***)).}
  \label{fig:Questionnaire1}
\vspace{-5mm}
\end{figure*}

\paragraph{Subjective perceived workload (NASA-TLX)}
\label{NASA}
The result showed that \textit{Mental Demand} was significantly lower in MetaRoundWorm than the Slide condition with a large effect size ($t(46) = -3.04$, $p < .01$, $d = -.88$ [large effect size]). Similarly, MetaRoundWorm demonstrated significantly reduced \textit{Temporal Demand} than the traditional teaching using slides ($U = 172.50$, $p < .05$, $r = .34$ [medium effect]). The levels of \textit{Frustration} were also significantly lower in MetaRoundWorm compared to the slide condition ($U = 143.00$, $p < .01$, $r = .43$ [medium effect]). No significant differences were observed in \textit{Physical Demand} ($p = .183$), \textit{Effort} ($U= 194.50$, $p = .051$, $r = .05$), and \textit{Performance} ($p = .073$). Figure~\ref{fig:Questionnaire1} (a) shows the Box plot results of the NASA-TLX.

\paragraph{AttrakDiff Questionnaire}
\label{AttrakDiff}

Independent samples t-tests revealed that MetaRoundWorm significantly outperformed the Slide condition in terms of Hedonic Quality - Stimulation (HQ-S) ($t(46)=4.58$, $p < .001$, $d=1.32$ [large effect size]), suggesting that MetaRoundWorm provided more stimulating experiences (Figure~\ref{fig:Questionnaire1} (b)), as corrected by Bonferroni ($\alpha$ = 0.0125), significant difference was found in Pragmatic Quality (PQ) ($t(46)= -2.14$, $p < .05$, $d=-.062$), however, no significant difference was found in Hedonic Quality - Identity (HQ-I) ($t(46)= -.26$, $p= .797$, $d= -.07$). Mann-Whitney U analysis results showed that MetaRoundWorm was also significantly better than the Slide condition in terms of attractiveness (ATT) ($U= 115.00$, $p< .001$, $ r = .52$ [large effect size]).

\paragraph{The Game Experience Questionnaire (GEQ)}


The MetaRoundWorm condition demonstrated significantly higher \textit{Sensory Imaginative Immersion} ($U = 443.50$, $p < .001$, $r = .46$ [large effect size]), \textit{Immersion Flow} ($t(46) = 4.23$, $p < .001$, $d= 1.22$ [large effect size]), \textit{Competence} ($t(46) = 3.36$, $p < .001$, $d= .97$ [large effect size]), \textit{Tension Annoyance} ($U = 147.00$, $p < .01$, $r = .42$ [large effect size]) and \textit{Positive Affect} ($U = 447.00$, $p < .001$, $r = .56$ [large effect size]). Conversely, the Slide condition showed significantly higher \textit{Negative Affect} ($U = 116.50$, $p < .001$, $r = .51$ [large effect size]). compared to the Slide condition. No significant differences were found for \textit{Challenge} ($t(46) = -1.53$,$p = .081$, $d=-.44$ [large effect size]) after Bonferroni correction ($\alpha$ = 0.007). See Figure~\ref{fig:Questionnaire1} (c) for details.


\paragraph{The Profile of Mood States- Short Form (POMS-SF)}
Analysis of POMS-SF data revealed that MetaRoundWorm participants experienced significantly lower negative mood states compared to the participants with slides across five dimensions: Tension–Anxiety ($t(38.71) = 2.87$, $p < .01$, $d= -.92$ [medium effect size]), Depression–Dejection ($t(34.23) = 2.79$, $p < .01$, $d=-.92$ [medium effect size]), Anger–Hostility ($t(32.73) = 2.94$, $p < .01$, $d=-1.01$ [medium effect size]), Fatigue–Inertia ( $t(38.34) = 5.33$, $p < .001$, $d=1.62$ [large effect size]), and Confusion–Bewilderment ($t(37.91) = 5.27$, $p < .001$, $d=1.64$ [large effect size]), with no difference in Vigor–Activity ($t(37.07)= -.91$, $p = .35$, $d=-.91$ ). The visualizations corroborate these findings, with Figure~\ref{fig:Questionnaire1} (d) showing consistently lower MetaRoundWorm compared to slide for negative dimensions, and boxplots demonstrating lower medians and narrower distributions in the MetaRoundWorm condition. Total Mood Disturbance was significantly lower in the MetaRoundWorm group ($t(37.50)=4.36$, $p < .001$, $d= -6.46$ [large effect size]), with the negative value indicating that positive mood exceeded negative mood, suggesting MetaRoundWorm exposure is more effective than traditional slide presentations in promoting positive emotional states.

\paragraph{Simulator Sickness Questionnaire (SSQ)}
\label{SSQ}
The SSQ assessment revealed moderate levels of discomfort among MetaRoundWorm users. Disorientation symptoms were most prominent, followed by oculomotor and nausea-related symptoms. Eye strain (41.7\%), fullness of head (37.5\%), general discomfort (33.3\%), and fatigue (33.3\%) were the most frequently reported symptoms. While 58.3\% of the participants experienced only minimal symptoms, 25.0\% reported severe symptoms, with 12.5\% reporting significant and 4.2\% concerning levels of discomfort. 

\subsection{Qualitative observation and feedback}

Through a systematic collection of observation records and participant feedback, this study qualitatively analyses the interactive behavioural patterns, learning strategies and affective performances of participants in the teaching process of MetaRoundWorm versus traditional slides. 

\paragraph{Interactive behavioural patterns} 
\label{sec:Interactive}
Participants under the MetaRoundWorm condition demonstrated notably different behavioral patterns compared to the traditional interactive slides condition. Almost all participants (N = 23) exhibited greater physical engagement when learning about roundworm infection pathways, actively using controllers for movement and interactive actions within the VR environment. In addition to this, participants who learned in the MetaRoundWorm condition were more willing to actively communicate with the experimenter and verbalize their feelings and opinions. In contrast, most of the participants in the traditional interactive slides condition (N = 14) maintained a single posture for the actions by clicking the screen throughout the learning process and remained largely silent.

The MetaRoundWorm condition allows users to explore virtual environments with a high degree of freedom, similar to the setting of open-world gaming, which encourages users' exploratory behavior. Specifically, they asked or expressed to the researcher: `\textit{Let me take a closer look at what the larvae look like,}' `\textit{I want to track this parasite to see where it migrates,}'  `\textit{Can I experience this process again to better understand immune evasion techniques?}'
In contrast, participants with the traditional approach of interactive slides demonstrated linear and structured patterns throughout their learning journey. In other words, participants (N = 17) repeatedly navigated between slides to compare information. It is worth mentioning that such behaviours are not found in the MetaRoundWorm condition, e.g., some participants (N = 7) requesting to take notes. 
Some participants (N = 8) who used traditional interactive slides reported frustration with the presentation of information, one participant commenting, `\textit{It was difficult to visualize three-dimensional structures from flat images...}' All participants (N = 24) found the interactive elements of the game highly engaging, appreciating the blend of entertainment and educational content. However, a few participants (N = 3) mentioned inconvenience in using controllers for navigation, expressing a preference for direct spatial walking controls within the virtual environment.

\paragraph{Learning strategies} 
\label{sec:strategies}
We observed that participants employed trial-and-error strategies in both the MetaRoundWorm and traditional interactive slides conditions. In the MetaRoundWorm, participants interacted with different virtual objects until they received correct feedback. In Level 6 (see Section~\ref{sec: Level 6}), where participants needed to select the appropriate immune cells or anti-parasitic medications to combat adult roundworms, we observed many participants (N = 20) continuously attempting different strategies until the adult roundworms were eliminated. One participant described, `\textit{I felt like a researcher... I like seeing the results immediately when I make mistakes.}' During the traditional interactive slides learning process, the participants experimented with different interactive features (N = 18), repeatedly attempting Assessment Modules and Visual Identification Functions (see Figure~\ref{fig:slide}) to obtain feedback on correct answers.

\paragraph{Affective performances} The two groups of participants exhibited different emotional responses. Most of the participants in the MetaRoundWorm group demonstrated high levels of curiosity and engagement, with nearly all participants (N = 23) describing the VR educational gaming experience as `\textit{very interesting}' and `\textit{fun}'. Participants in the MetaRoundWorm group also showed a stronger `\textit{sharing impulse}', all participants (N = 24) actively describing their experiences and explaining the content they learned to researchers after the experiment. However, some participants (N = 7) noted that VR made them excited, occasionally distracting from learning, such as, `\textit{I was so focused on animations and games that I may have missed some information [knowledge]}'. In contrast, participants in the traditional interactive slides group generated markedly different emotional responses. Most of the participants in this group (N = 16) described their emotional state as `\textit{bored}' rather than `\textit{excited}'. One participant said: `\textit{These interactions made me feel like I was answering test questions}'.

\section{Discussion}

\subsection{Learning Effectiveness of MetaRoundWorm}

Our results indicate that MetaRoundWorm exhibited significant learning advantages over traditional interactive slides in immediate post-test assessments, consistent with different cognitive levels of Bloom's Taxonomy, with enhanced knowledge acquisition at the level of memory and comprehension, supporting \textbf{[H1]}. This advantage can be attributed to several factors. First, MetaRoundWorm provided an immersive experience that activated learning interest~\cite{gao2018learning}. Furthermore, in MetaRoundWorm, participants not only passively observed different stages of the lifecycle of the parasite but actively ``follow'' the parasite into various human organs (see Section~\ref{sec:Interactive}), in line with \textbf{[H6]}. This kind of active exploration behavior may lead to higher SSQ scores (see Section~\ref{SSQ} for details). This interactive engagement potentially enhanced understanding of complex biological processes~\cite{reinke2021immersive}. The VR environment also supported active exploration, encouraging deeper knowledge acquisition by allowing participants to freely navigate and select viewing perspectives. This aligns with constructivist learning theory~\cite{roussos1999learning}, promoting active construction of knowledge frameworks rather than passive information reception. Additionally, the gamified elements of MetaRoundWorm significantly enhanced hedonic quality -- stimulation and attractiveness (see Section~\ref{AttrakDiff}). Complex immunological concepts were transformed into tangible multisensory experiences involving visual, auditory, and interactive elements, making the learning process engaging. Although the Slide condition contained interactive elements, it contained less gamified design. For complex scientific concepts, gamification elements can enhance learning motivation, consistent with the study by Falah et al.~\cite{falah2021identifying}. Moreover, MetaRoundWorm reduced participants' subjective perceived workload (see Section~\ref{NASA}).  According to NASA-TLX results, MetaRoundWorm elicited significantly lower \textit{Mental Demand} scores compared to traditional interactive slides. This violates \textbf{[H5]}, The possible reasons are with slide presentations, text-heavy content and realistic images of roundworms may induce cognitive overload, negatively affecting learning outcomes and attentional allocation~\cite{SWELLER1988257}. A participant in the Interactive Slide group mentioned, ``\textit{I am threatened by the worms}''. In contrast, in MetaRoundWorm, this effect was mitigated, with no participants reporting scared feelings to the researcher during gameplay in MetaRoundWorm.

The follow-up test revealed better knowledge retention in the MetaRoundWorm group, supporting \textbf{[H2]}, which suggests deeper learning at both the application and analysis levels in Bloom's Taxonomy. Even after one week, the VR group maintained significantly higher scores than the Slides group. This memory retention advantage may be related to the contextual memory provided by the VR environment~\cite{smith2019virtual}, as one participant mentioned:``\textit{I remember seeing roundworm eggs in the mouth.}'' Immersive spatial navigation experiences strengthened spatial memory associations, creating richer mnemonic cues and enhancing subsequent information retrieval~\cite{tuena2024bodily}.

Interestingly, correlation analyses revealed that longer learning time was associated with lower test performance in the MetaRoundWorm group, while the opposite pattern was observed in the Interactive Slides condition. This suggests that immersive environments such as MetaRoundWorm may enable more efficient learning, where key concepts are acquired in a shorter time. In contrast, slide-based learning may rely more on repeated exposure and longer participation. The greater variability in the MetaRoundWorm group, especially in follow-up scores, may reflect differences in individual exploration strategies (see Section~\ref{sec:Interactive} \& Section~\ref{sec:strategies}). These findings highlight the need to balance user autonomy and guided learning in immersive environments to support consistent learning outcomes.

\subsection{Emotional Response and Motivation}

The POMS-SF results indicate that the short-term negative affect change in the MetaRoundWorm group was significantly lower than in the interactive slide group, supporting \textbf{[H3]}. Emotional fluctuations can affect learning outcomes, as a positive emotional state can enhance cognition and promote creative thinking as well as problem-solving abilities~\cite{isen2004some}. Moreover, in the MetaRoundWorm condition, participants reported stronger \textit{Positive Affect} and better scores in \textit{Attractiveness} and \textit{Hedonic Quality – Stimulation}. These results also support \textbf{[H4]}. These positive emotions may directly facilitate attention allocation and the process of knowledge retention, according to the theory of Tyng et al.~\cite{tyng2017influences}, emotion has a significant impact on attention, memory, and learning effectiveness. In particular, the GEQ \textit{Competence} Subscale score was also significantly higher for MetaRoundWorm compared to the interactive Slide group. As \textit{Competence} is positively correlated with self-efficacy, a higher self-efficacy is likely to foster a more active participation in learning tasks~\cite{gist1992self}.

\subsection{Design Implications for Educational VR Game}

Based on our findings, we propose the following design implications (DIs) for future educational VR games, particularly in the fields of medicine, biology, and epidemiology. \textbf{DI1}: Educational VR should integrate immersive storytelling and clear task-centered mechanics to contextualize complex scientific content effectively. Allowing learners to actively explore virtual scenarios facilitates deeper conceptual understanding compared to linear, passive learning methods. \textbf{DI2}: Using multimodal interactions (visual, auditory, and haptic feedback) to improve conceptual clarity and knowledge retention. Embodied interactions in VR environments provide rich memory-aiding cues that provide immediate and meaningful feedback to support the learner's memory formation process. \textbf{DI3}: Designers should focus on the learner's emotional experience to maximize engagement and minimize negative impact, especially in sensitive topics. Educate learners to feel empowered and confident, engage more actively in learning tasks, and improve overall motivation for learning. \textbf{DI4}:  Design intuitive interactions to minimize cognitive load and physical discomfort. Simple, natural controls improve user experience and maintain immersion, thus ensuring sustained cognitive engagement. Incorporating these design principles into future educational VR applications will likely result in higher efficacy in teaching complex biomedical concepts.

\subsection{Limitations and Future Work}



Although this research showcases several educational advantages of MetaRoundWorm, it has some limitations, as follows. The current game design encompasses the four cognitive levels of Bloom's Taxonomy (Remember through Evaluate), covering the understanding of parasite life cycles, the application of immune mechanisms, and the use of treatment strategies. Our current game design does not include Evaluate-Level because we set up knowledge tests before and after the experiment; however, for future game testing products that are intended to be used commercially, there is a need to incorporate settings for Evaluate-Level into the game design (e.g., setting up a simple in-game test to assess the player's knowledge) in order to test the validity of the education. We can also incorporate Create-level activities (e.g., allowing players to design prevention strategies) to enhance cognitive engagement in parasitology education. Participants were limited to a particular demographic characteristic, e.g., Asian post-secondary students, who likely exhibited elevated technological acceptance and a strong willingness to learn. This constrains the generalizability of our results to a broader community, particularly among learners unfamiliar with the technology or demographics with significant age disparities. Future research should enhance the diversity of samples (e.g.,K-12 students, non-student populations, and even tech-savvy learners) to validate the prevalence of the impact of virtual reality education. The moderate disorientation (41.7\% eye strain, 37.5\% “fullness of head”) observed in the SSQ results may affect learning outcomes. This may be related to factors such as the color scheme, movement patterns, and head rotation in the game. Although we used the teleport movement method to reduce player discomfort, future research should focus on minimizing excessive player movement during prolonged learning sessions and avoiding overly vibrant visuals of the game (e.g. colour tones, contrast of text and background) that may increase eye strain~\cite{li2025visual}. We will also extend our studies (e.g., 3 months, 6 months, or longer) to evaluate the long-term retention of information and conceptual understanding resulting from VR educational experiences. Next, we will focus particularly on changes in knowledge organisation, rather than only on memory retention. Additionally, we will analyze the educational impacts of VR across many disciplines (e.g., physics, chemistry, anatomy) to determine the most effective application domains and content characteristics for VR education. We urge more study to determine which kinds of knowledge are most effectively conveyed via VR and which are better suited for traditional techniques.

\section{Conclusion}
This study presents \textit{MetaRoundWorm}, a VR-based escape room game designed to enhance education on parasitic infections and immune responses through immersive and gamified learning. leverages the metaverse to show the micro-world of the immune system responding to the roundworm. Our results demonstrate that \textit{MetaRoundWorm} significantly improves immediate learning outcomes, enhances user engagement, and fosters positive emotional responses compared to traditional interactive slide-based teaching methods. Participants not only acquired knowledge more effectively but also retained it over time, indicating the long-term educational potential of immersive VR environments.
Furthermore, \textit{MetaRoundWorm} led to lower perceived workload and negative emotional states, while promoting higher stimulation, attractiveness, and a sense of competence. This study highlights the potential of VR games as an innovative educational method, effectively transforming complex biomedical education into an engaging experience.


\acknowledgments{%
This research was supported by the Hong Kong Polytechnic University's Start-up Fund for New Recruits (No. P0046056), Departmental General Research Fund (DGRF) from HK PolyU ISE (No. P0056354), and PolyU RIAM -- Research Institute for Advanced Manufacturing (No. P0056767). Xian Wang was supported by a grant from the PolyU Research Committee under student account codes RMHD.
}

\newpage
\bibliographystyle{abbrv-doi}
\bibliography{main}
\end{document}